\documentclass[epjCONF]{svjour}

\usepackage{graphics}
\usepackage[varg]{txfonts}
\usepackage[latin1]{inputenc}

\session-title{Conference Title, to be filled}

\begin{document}

\title{KOI-3158: The oldest known system of terrestrial-size planets}

\author{T.~L.~Campante\inst{1,2}\fnmsep\thanks{\email{campante@bison.ph.bham.ac.uk}}
\and T.~Barclay\inst{3,4}
\and J.~J.~Swift\inst{5}
\and D.~Huber\inst{3,6,7}
\and V.~Zh.~Adibekyan\inst{8,9}
\and W.~Cochran\inst{10}
\and C.~J.~Burke\inst{6,3}
\and H.~Isaacson\inst{11}
\and E.~V.~Quintana\inst{6,3}
\and G.~R.~Davies\inst{1,2}
\and V.~Silva Aguirre\inst{2}
\and D.~Ragozzine\inst{12}
\and R.~Riddle\inst{13}
\and C.~Baranec\inst{14}
\and S.~Basu\inst{15}
\and W.~J.~Chaplin\inst{1,2}
\and J.~Christensen-Dalsgaard\inst{2}
\and T.~S.~Metcalfe\inst{16,2}
\and T.~R.~Bedding\inst{7,2} 
\and R.~Handberg\inst{1,2}
\and D.~Stello\inst{7,2}
\and J.~M.~Brewer\inst{17}
\and S.~Hekker\inst{18,2}
\and C.~Karoff\inst{2,19}
\and R.~Kolbl\inst{11}
\and N.~M.~Law\inst{20}
\and M.~Lundkvist\inst{2}
\and A.~Miglio\inst{1,2}
\and J.~F.~Rowe\inst{3,6}
\and N.~C.~Santos\inst{8,9,21}
\and C.~Van Laerhoven\inst{22}
\and T.~Arentoft\inst{2}
\and Y.~P.~Elsworth\inst{1,2}
\and D.~A.~Fischer\inst{17}
\and S.~D.~Kawaler\inst{23}
\and H.~Kjeldsen\inst{2}
\and M.~N.~Lund\inst{2}
\and G.~W.~Marcy\inst{11}
\and S.~G.~Sousa\inst{8,9,21}
\and A.~Sozzetti\inst{24}
\and T.~R.~White\inst{25}
}
\institute{School of Physics and Astronomy, University of Birmingham, Edgbaston, Birmingham, B15 2TT, UK
\and Stellar Astrophysics Centre (SAC), Department of Physics and Astronomy, Aarhus University, Ny Munkegade 120, DK-8000 Aarhus C, Denmark 
\and NASA Ames Research Center, Moffett Field, CA 94035, USA
\and Bay Area Environmental Research Institute, 596 1st Street West, Sonoma, CA 95476, USA
\and Department of Astronomy and Department of Planetary Science, California Institute of Technology, MC 249-17, Pasadena, CA 91125, USA
\and SETI Institute, 189 Bernardo Avenue \#100, Mountain View, CA 94043, USA
\and Sydney Institute for Astronomy, School of Physics, University of Sydney, Sydney, Australia
\and Centro de Astrof\'isica, Universidade do Porto, Rua das Estrelas, 4150-762 Porto, Portugal
\and Instituto de Astrof\'isica e Ci\^encias do Espa\c{c}o, Universidade do Porto, Rua das Estrelas, 4150-762 Porto, Portugal
\and Department of Astronomy and McDonald Observatory, The University of Texas at Austin, USA
\and Astronomy Department, University of California, Berkeley, CA 94720, USA
\and Department of Physics and Space Sciences, Florida Institute of Technology, 150 West University Boulevard, Melbourne, FL 32901, USA
\and Division of Physics, Mathematics, and Astronomy, California Institute of Technology, Pasadena, CA 91125, USA   
\and Institute for Astronomy, University of Hawai`i at M\=anoa, Hilo, HI 96720-2700, USA
\and Department of Astronomy, Yale University, New Haven, CT 06520, USA
\and Space Science Institute, Boulder, CO 80301, USA
\and Department of Physics, Yale University, New Haven, CT 06511, USA
\and Max Planck Institute for Solar System Research, G\"ottingen, Germany
\and Department of Geoscience, Aarhus University, H{\o}egh-Guldbergs Gade 2, DK-8000 Aarhus C, Denmark
\and Department of Physics and Astronomy, University of North Carolina at Chapel Hill, Chapel Hill, NC 27599-3255, USA
\and Departamento de F\'isica e Astronomia, Faculdade de Ci\^encias, Universidade do Porto, Rua do Campo Alegre, 4169-007 Porto, Portugal
\and Department of Planetary Sciences, University of Arizona, 1629 E University Blvd., Tucson, AZ 85721, USA
\and Department of Physics and Astronomy, Iowa State University, Ames, IA 50011, USA
\and INAF -- Osservatorio Astrofisico di Torino, Via Osservatorio 20, I-10025 Pino Torinese, Italy
\and Institut fur Astrophysik, Georg-August-Universit\"at G\"ottingen, Friedrich-Hund-Platz 1, 37077 G\"ottingen, Germany
}

\abstract{
The first discoveries of exoplanets around Sun-like stars have fueled efforts to find ever smaller worlds evocative of Earth and other terrestrial planets in the Solar System. While gas-giant planets appear to form preferentially around metal-rich stars, small planets (with radii less than four Earth radii) can form under a wide range of metallicities. This implies that small, including Earth-size, planets may have readily formed at earlier epochs in the Universe's history when metals were far less abundant. We report {\it Kepler} spacecraft observations of KOI-3158, a metal-poor Sun-like star from the old population of the Galactic thick disk, which hosts five planets with sizes between Mercury and Venus. We used asteroseismology to directly measure a precise age of $11.2\pm1.0\:{\rm Gyr}$ for the host star, indicating that KOI-3158 formed when the Universe was less than $20\,\%$ of its current age and making it the oldest known system of terrestrial-size planets. We thus show that Earth-size planets have formed throughout most of the Universe's $13.8$-billion-year history, providing scope for the existence of ancient life in the Galaxy.
}

\maketitle

\section{Introduction}\label{sec:intro}
The NASA {\it Kepler} spacecraft \cite{KeplerMission} uses precise photometry to measure the periodic dips in starlight due to transits of planets across the face of their stars. {\it Kepler} has so far successfully detected over $4{,}000$ exoplanet candidates using the transit method, of which approximately $40\,\%$ are in multiple-planet systems \cite{BurkeCandidates4}. Transit-like signals indicative of five planets were detected by {\it Kepler} over the course of four years of nearly continuous observations of KOI-3158 (also known as HIP~94931 and KIC~6278762).

KOI-3158 is a cool main-sequence star (spectral type K0V) with apparent magnitude $V\!=\!8.86$ and lying at a distance of $36\:{\rm pc}$ \cite{Hipparcos} in the constellation Lyra, making it the brightest and closest multiple-planet host detected by {\it Kepler}. It is a high-proper-motion star \cite{Hipparcos}, with an annual motion across the celestial sphere in excess of $0.5\:{\rm arcsec}$. A high-resolution spectrum obtained with the Keck/HIRES spectrograph showed that KOI-3158 has a paucity of heavy elements with only about $30\,\%$ of the Sun's iron abundance. Furthermore, it is overabundant in $\alpha$-process elements (such as silicon and titanium) for a star of its metallicity.

\section{Hierarchical triple system}\label{sec:triple}
During spectroscopic observations with the Keck telescope, a fainter companion was visually detected on the HIRES guide camera at an angular distance of $1.8\:{\rm arcsec}$, thus being unresolved in {\it Kepler} observations. Follow-up observations using the Robo-AO system \cite{RoboAO} on the Palomar 60-inch telescope were used to determine the level of resulting contamination in the {\it Kepler} bandpass. The two components are co-moving as implied by their systemic radial velocity, with the orbital period of the secondary around the primary estimated to be $\sim\!430\:{\rm yr}$. Cross-correlating the spectrum of the secondary with that of a template M dwarf revealed that it in fact comprises two M dwarfs. Therefore, KOI-3158 is the primary star in a hierarchical triple system.

\section{A member of the thick disk}\label{sec:thick}
An analysis of its chemical properties and kinematics revealed that KOI-3158 belongs to the Galactic thick disk. Moreover, KOI-3158 is a member \cite{AF06} of the Arcturus stellar stream, a moving group of stars from the thick disk. The origin of the Arcturus stream has been the matter of debate \cite{Klement10}. Initially thought to be of extragalactic origin, as a result of a merger event, it has more recently been interpreted as arising from dynamical perturbations within the Galaxy.

\section{Asteroseismic analysis}\label{sec:seis}
\begin{figure}
\centering
\resizebox{\columnwidth}{!}{\includegraphics{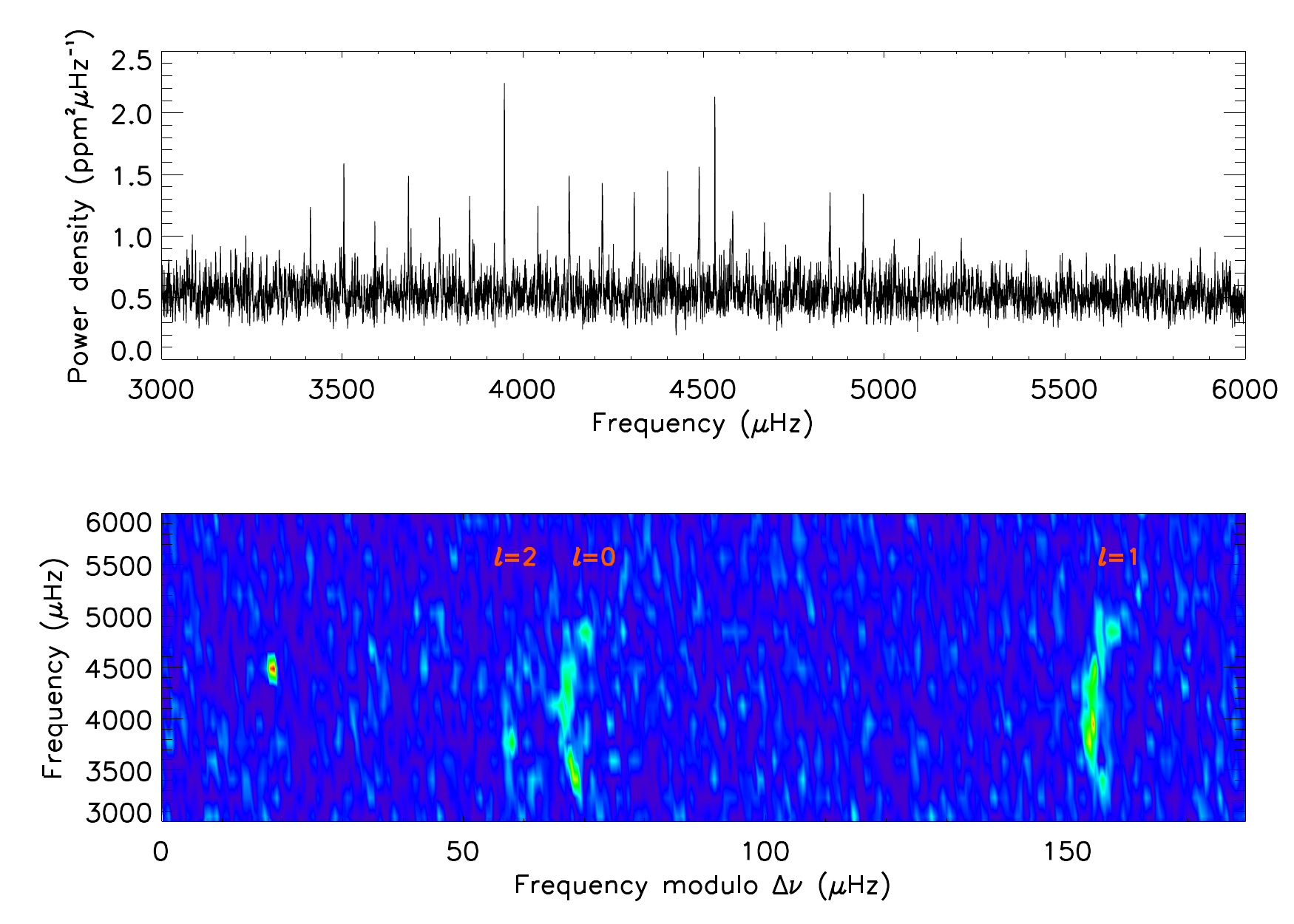}}
\caption{Top panel: Frequency-power spectrum of the flux time series of KOI-3158 over the frequency range occupied by the solar-like oscillations. Bottom panel: Frequency-power spectrum in \'echelle format. This is the graphical equivalent to slicing the spectrum into segments of length $\Delta\nu$ (i.e., the large frequency separation) and stacking them one on top of the other. Note that, in order to center the power ridges on the diagram, frequencies have been shifted sideways by subtracting a fixed reference. Power ridges are tagged according to mode degree, $l$.}\label{fig:PS}
\end{figure}

We detected solar-like oscillations \cite{JCDReview} in the flux time series of KOI-3158 (Fig.~\ref{fig:PS}). Stellar properties were estimated by matching asteroseismic and spectroscopic observables to grids of stellar evolutionary models \cite{HuberEnsembleKOI}. We measured a stellar mass of $0.758\pm0.043\,{\rm M}_\odot$ and a radius of $0.752\pm0.014\,{\rm R}_\odot$. Further analysis based on individual oscillation frequencies \cite{VSA13} revealed that the star, and hence the planetary system, has an age of $11.2\pm1.0\:{\rm Gyr}$, commensurate with those of thick-disk field stars \cite{Reddy06}.

\section{Characterization of the planetary system}\label{sec:planetary}
\subsection{Transit analysis}\label{sec:transit}
We fitted a five-planet transit model to the {\it Kepler} data using an affine-invariant Markov chain Monte Carlo algorithm \cite{Foreman-Mackey}, having fixed the dilution of the light curve at $3.94\,\%$ based on our adaptive optics observations. The precision with which we measured the planetary radii varies between $2.9\,\%$ and $5.5\,\%$. KOI-3158.01 is the innermost and smallest planet (within $2\sigma$ of the size of Mercury). Its radius was measured with a precision of $\sim\!100\:{\rm km}$. All five planets are sub-Earth-sized with monotonically increasing radii as a function of orbital distance: $0.403$, $0.497$, $0.530$, $0.546$, and $0.741\,R_\oplus$. KOI-3158.02, KOI-3158.03, and KOI-3158.04 have very similar radii, respectively within $2\sigma$, $1\sigma$, and $1\sigma$ of the size of Mars. Finally, KOI-3158.05 has a size intermediate to Mars and Venus. KOI-3158 thus expands the population of planets found in low-metallicity environments from the mini-Neptunes around Galactic halo's Kapteyn's star \cite{Kapteyn} down to the regime of terrestrial-size planets.

This system is highly compact, both in terms of its architecture and in a dynamical sense. All planets orbit the parent star in less than 10 days, or within $0.08\:{\rm AU}$, roughly one-fifth the size of Mercury's orbit. Furthermore, all adjacent planet pairs are close to being in exact orbital resonances, with period ratios no more than $\sim\!2\,\%$ in excess of strong $5\colon\!4$, $4\colon\!3$, $5\colon\!4$, and $5\colon\!4$ first-order mean-motion resonances (MMRs) as one moves toward the outermost pair.

\subsection{System validation}\label{sec:validation}
Several lines of evidence validate this system as a true five-planet system orbiting the target star. We invoke statistical arguments \cite{Lissauer12} to exclude chance-alignment blends (such as background eclipsing binaries) as the cause of one or more transit-like signals. The only plausible false-positive configuration of four planets and one background eclipsing binary is rejected at the $99.9\,\%$ level. Pairs of planets within {\it Kepler}'s multiple systems show a slight propensity to be in or near MMRs \cite{Lissauer11}, a feature also predicted by theoretical models of planet formation and migration. We thus invoke the non-randomness of the observed multi-resonant chain to assert, at the $90\,\%$ level, that all five planets form a single system orbiting the same star. Finally, we establish that the planets transit the target star and not one of its M-dwarf companions. This was achieved by testing the dynamical stability of the five planets if they were to all orbit either of the M-dwarf companions. Planetary masses were estimated for an exhaustive range of possible compositional schemes using theoretical mass-radius relations \cite{Fortney07,Lopez12}. Such a system would become unstable within $10^2$ to several $10^3$ years for all compositional schemes tested.

\section{Concluding remarks}\label{sec:conclusions}
At $11.2\pm1.0\:{\rm Gyr}$, KOI-3158 is the oldest known system of terrestrial-size planets. The precision with which the age of KOI-3158 has been determined from asteroseismology ($\sim\!9\,\%$) is an impressive technical achievement that was only made possible due to the extended and high-quality photometry provided by the {\it Kepler} mission. We show that Earth-size planets have formed throughout most of the Universe's 13.8-billion-year history, leaving open the possibility for the existence of ancient life in the Galaxy. Remarkably, by the time Earth formed, this star and its Earth-size companions were already older than our planet is today.

\end{document}